\documentclass{article}

\usepackage{PRIMEarxiv}

\usepackage[utf8]{inputenc} 
\usepackage[T1]{fontenc}    
\usepackage{hyperref}       
\usepackage{url}            
\usepackage{booktabs}       
\usepackage{amsfonts}       
\usepackage{nicefrac}       
\usepackage{microtype}      
\usepackage{lipsum}
\usepackage{fancyhdr}       
\usepackage{graphicx}       
\graphicspath{{media/}}     

\title{CSDR-BERT: a pre-trained scientific dataset match model for Chinese Scientific Dataset Retrieval}

\author{
  Xintao Chu \\
  College of Computer Science and Engineering \\
  North Minzu University \\
  Yinchuan\\
  \texttt{xtchu@stu.nun.edu.cn} \\
   \And
  Jianping Liu \\
  College of Computer Science and Engineering \\
  North Minzu University \\
  Yinchuan\\
  \texttt{liujianping01@nmu.edu.cn} \\
   \And
  Jian Wang \\
  Agricultural Information Institute \\
  Chinese Academy of Agricultural Sciences \\
  Beijing\\
  \texttt{wangjian01@caas.cn} \\
   \And
  Xiaofeng Wang \\
  College of Computer Science and Engineering \\
  North Minzu University \\
  Yinchuan\\
  \texttt{xfwang@nmu.edu.cn} \\
   \And
  Yingfei Wang \\
  College of Computer Science and Engineering \\
  North Minzu University \\
  Yinchuan\\
  \texttt{20217426@stu.nmu.edu.cn} \\
    \And
  Meng Wang \\
  College of Computer Science and Engineering \\
  North Minzu University \\
  Yinchuan\\
  \texttt{20217441@stu.nmu.edu.cn} \\
      \And
  Xunxun Gu \\
  College of Computer Science and Engineering \\
  North Minzu University \\
  Yinchuan\\
  \texttt{3043691838@qq.com} \\
}

\begin{document}
\maketitle

\begin{abstract}
As the number of open and shared scientific datasets on the Internet increases under the open science movement, efficiently retrieving these datasets is a crucial task in information retrieval (IR) research. In recent years, the development of large models, particularly the pre-training and fine-tuning paradigm, which involves pre-training on large models and fine-tuning on downstream tasks, has provided new solutions for IR match tasks. In this study, we use the original BERT token in the embedding layer, improve the Sentence-BERT model structure in the model layer by introducing the SimCSE and K-Nearest Neighbors method, and use the cosent loss function in the optimization phase to optimize the target output. Our experimental results show that our model outperforms other competing models on both public and self-built datasets through comparative experiments and ablation implementations. This study explores and validates the feasibility and efficiency of pre-training techniques for semantic retrieval of Chinese scientific datasets.
\end{abstract}
\setlength{\baselineskip}{15.7pt}
\section{Introduction}
In the field of scientific research, mining and analyzing scientific data is a crucial task. Scientific data is often organized in the form of datasets, and the retrieval of these datasets is a vital function in many scientific data warehouses \cite{122}. With the rapid growth of scientific datasets, traditional retrieval methods are unable to meet the demands of researchers for fast and accurate answers. Recently, the use of pre-training models and text matching computation is an effective solution to these problems. Semantic text matching is a crucial method for text data analysis, commonly used for text clustering and classification, and is central to achieving scientific data retrieval \cite{chen2021long}. Therefore, it is important to study and improve semantic text matching models.

Semantic text matching aims to determine whether pairs of sentences have the same meaning \cite{zhao2020interactive}. Determining the similarity between sentence pairs is a critical task and is widely used in information retrieval, machine translation, question and answer, and recommendation systems, among others \cite{zhang2020enhanced}. With the introduction of the Transformers framework, pre-training models have seen rapid development. The current mainstream approach for semantic text matching is to fine-tune a downstream task using pre-training models (e.g. Sentence-BERT). Pre-training models can provide a better starting point for the downstream task to learn, allowing for faster convergence and improved performance \cite{zheng2020pre}. However, a key challenge in text matching in the field of scientific data is that terms from different research fields are not well established in databases, and texts with the same meaning can have different expressions (e.g. chickenpox and varicella). Furthermore, different research fields often have unique expressions for their terms (e.g. glucose and C6H12O6).

To address these issues, we propose a semantic text matching model based on pre-training, which enhances the representation of sentences through contrast learning. To improve text matching in the scientific data domain, we use the Chinese Scientific Literature (CSL) dataset, which is the first publicly available scientific dataset obtained from the National Engineering and Technology Research Center for Science and Technology Resources Sharing Services \cite{li2022csl}. Considering the unique characteristics of scientific data texts, we construct CSDR-BERT, a scientific dataset text matching model that combines pre-training, contrast learning, and KNN, to enhance the accuracy of scientific dataset text similarity judgment.

The main contributions of this paper are as follows:1)We collected a scientific dataset for semantic text matching and created word lists.2)We improved the structure of the Sentence-BERT model to develop a semantic retrieval model, CSDR-BERT, for Chinese scientific datasets. The model improves text representation by using contrast learning to build clusters for semantic text matching tasks.3)We use SimCSE and CSL datasets for pre-training the models, thereby enhancing the knowledge base of the pre-trained models and improving their semantic matching abilities.4)We conducted experiments on both self-constructed and public datasets, and the results verified the effectiveness of the model.

\section{Related work}
\subsection{Text matching}
Calculating semantic similarity is a crucial task in text matching models. Early methods would extract keywords from text objects and then calculate their cosine distances \cite{li2013distance}. However, the emergence and widespread use of deep learning models has greatly contributed to the advancement of natural language processing due to their ability to adapt to multiple levels of natural language and its intrinsic logic \cite{li2021word}. From the earliest Deep Neural Network (DNN) that used feedforward neural networks to map text sequences, to Convolutional Neural Networks (CNN) which share parameters in a fixed-size sliding window, and then to Recurrent Neural Network (RNN) to capture long-term dependencies in text sequences \cite{liang2017text}. Long and Short Memory Neural Network (LSTM) is a specialized form of RNN that aims to solve the long-distance dependency problem. 

Text representation is the core of text matching \cite{zhu2015research}. Early work was inspired by siamese network architectures, which use separate neural networks to encode the two input sequences into high-level representations. Huang et al. \cite{huang2013learning} proposed DSSM, which uses DNN to represent queries and headings as low-dimensional semantic vectors to handle the large number of words commonly found in such tasks. Gia-Hung Nguyen et al. \cite{nguyen2017dsrim} propose DSRIM, which models the relational semantics of text at the raw data level. The C-DSSM model proposed by Yelong Shen et al. \cite{shen2014learning} extracts local contextual textual representations from the text using convolutional neural networks. In the same year, Yelong Shen et al. \cite{shen2014latent} used sliding windows to convert text to Letter-trigram form to capture more contextual information. Baotian Hu et al. \cite{hu2014convolutional} proposed ARC-1, which obtains multiple combinatorial relationships between adjacent feature maps through convolutional layers of different terms. Sunil Mohan et al. \cite{mohan2018fast} propose Delta, a model that uses the Word2Vector method for embedding text. H. Palangi et al. \cite{palangi2014semantic}  propose LSTM-DSSM, which treats words in the text as sequences of words and uses LSTM to capture contextual semantics. Song et al. \cite{song2018deep} use the gated cyclic unit BI-GRU to extract rich contextual features on a sentence-by-sentence basis and represent word-level and sentence-level interaction information through an attention mechanism. Early on, LSTM and its variants have been widely used for semantic text matching and have achieved good performance. 

Recent work has shown that pre-training models on large corpora can learn generic language representations, which can be beneficial for downstream NLP tasks and can avoid the need to train new models from scratch.
\subsection{Pre-training techniques and their application in text matching}
\subsubsection{Pre-training model}
Due to the high cost of annotation, building large-scale tag datasets for most NLP tasks is a significant challenge. In contrast, large-scale unlabeled corpora are relatively easy to construct. By using these corpora, models can learn good representations from them and apply these representations to other tasks. In the early stage, words were represented as dense vectors for pre-training word embeddings. Word2vec \cite{mikolov2013efficient} is a typical example of this, where it is used for different NLP tasks by pre-training word embeddings. Since most NLP tasks go beyond the word level, higher-level pre-trained neural encoders such as sentences have emerged. After the Transformer was proposed in 2017, pre-trained models have demonstrated their power in learning generic language representations, such as BERT \cite{devlin2018bert}, GPT \cite{radford2018improving}, XLNet \cite{yang2019xlnet}, T5 \cite{raffel2020exploring}, etc. Since the introduction of the BERT family, fine-tuning has become the dominant method for adapting pre-trained models to downstream tasks.
\subsubsection{Comparative learning}
Recently, a contrastive learning framework has been widely used in self-supervised tasks. It can use unlabeled datasets to enhance the potential semantic representation. Nils Reimers et al. \cite{reimers2019sentence} propose Sentence-BERT, which uses siamese neural networks to generate embedding vectors of sentences with semantic meaning. Two sentences from the same paragraph are considered positive examples, otherwise, they are considered negative examples. Fang et al. \cite{fang2020cert} propose the CERT method, which creates two different but semantically similar sentences as positive examples using back-translation. CERT uses BERT as its encoder and InfoNCE as the contrast loss function. Gao et al. \cite{gao2021simcse} proposed the SimCSE method, which generates positive examples of the model by passing a sentence through two different dropout layers, with the rest of the sentences in the same batch being negative examples.
\subsubsection{Research on text matching based on pre-trained models}
Pre-trained language models, represented by BERT, are also widely used in text matching tasks. These models are first pre-trained on a large corpus and then fine-tuned for a specific text matching task to learn domain-specific knowledge. Sneha Choudhary et al. \cite{choudhary2020document} propose the use of BERT fused with traditional methods (TF-IDF, BM25, etc.) to address the inability of bag-of-words methods to capture the semantics of context by creating semantic-rich document embeddings through BERT. The limitations of Term Frequency Inverse Document Frequency (TF-IDF) are likewise compensated for by combining contextual embeddings. Andre Esteva et al. \cite{esteva2021covid} propose a text enhancement technique that divides a document into pairs of paragraphs and the citations contained therein, creating a large number of (citation headings, paragraph) tuples for training the retrieval module. Gianluca Moro et al. \cite{moro2021efficient} proposed a self-supervised method SUBLIMER. This method does not use labels but creates a potential space in an unlabeled corpus of papers, and the distance in the space is a measure of semantic similarity. The core idea of creating potential space is to use the references between papers to define the positive or negative correlation between them. Sohrab Ferdowsi et al. \cite{teodoro2021information} used a deviant DFR model with randomness and a Dirichlet a priori smoothing-based LMD model for language retrieval to enhance text retrieval matching.
\subsection{Summary}
In summary, approaches based on pre-trained models have become the mainstream of semantic text matching model research and have achieved good results. The problem of semantic matching for Chinese scientific datasets is a typical example of NLP semantic matching problems. There is a lack of pre-training models for Chinese scientific datasets and related research on semantic text matching. Therefore, this paper will utilize the latest results on pre-training models to solve the retrieval match problem for Chinese scientific datasets.
\section{Method}
\subsection{Improving the Sentence-BERT model for retrieval matching}
\begin{figure}[!h]
  \centering
  \includegraphics[width=10cm,height=8cm]{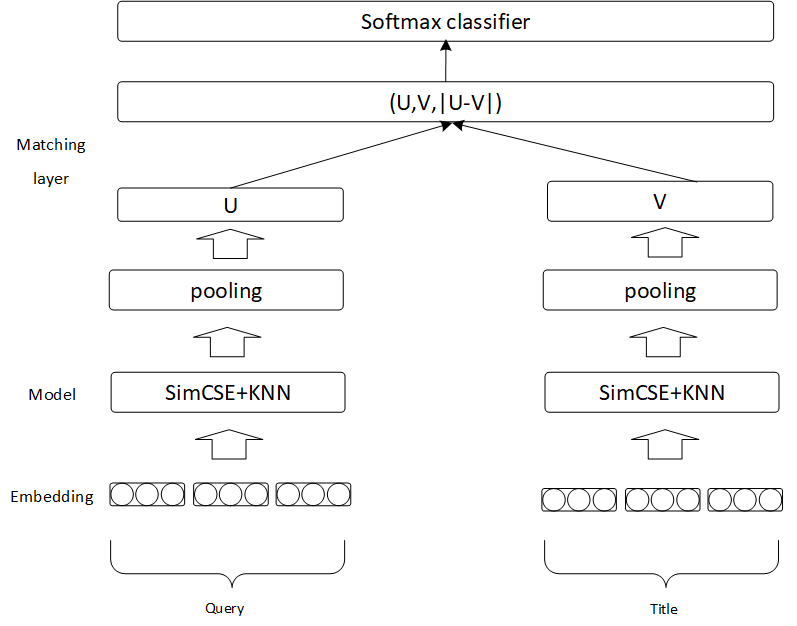}
  \caption{Improved Sentence-BERT model for retrieval matching}
  \label{fig:f1}
\end{figure}

According to the characteristics of retrieval, we chose Sentence-BERT as the basic method and improved it to adapt to the retrieval matching task. We replaced the pre-training model BERT with the SimCSE pre-training model and added the KNN algorithm, changed the loss function to the Cosent function, and verified the effectiveness of the model through iterative trials. The structure of the proposed model is shown in Figure \ref{fig:f1}
\subsubsection{Embedding layer}
\begin{figure}[!htbp]
  \centering
  \includegraphics[width=13cm,height=5cm]{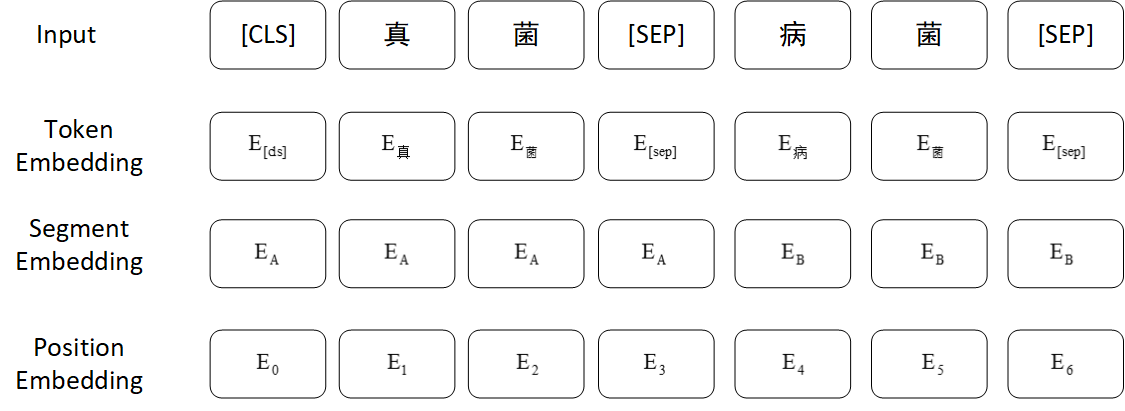}
  \caption{Embedded layer}
  \label{fig:f2}
\end{figure}

BERT takes each word (token) in the input text and feeds it into a token embedding layer, which maps each word to a low-dimensional vector space and transforms it into a text representation vector. The embedding process consists of three components: 1) Token embedding, which converts words into a uniform dimension. 2) Segment embedding, which enables the model to distinguish between two sentences of text. 3) Position embedding, which enables the model to understand the order of words and word order.As shown in figure \ref{fig:f2}
\subsubsection{Model layer}
(1) Unsupervised SimCSE comparative learning framework

The contrast learning framework is widely used in self-supervised tasks. Contrast learning is a similarity-based training strategy that aims to bring positive samples closer together and negative samples further apart. In the unsupervised SimCSE method, after a sentence in the data sample is encoded to generate a sentence embedding, noise is added through the dropout technique to obtain two different sentence vectors to construct positive samples, while the other sentence embeddings in the sample are used as negative samples, as a way to better learn the data representation information. The specific formula is as follows.
\begin{equation}
{l_{\rm{i}}} =  - \log \frac{{{e^{sim({h_i},h_i^ + )/t}}}}{{\sum {_{j = 1}^N{e^{{\mathop{\rm sim}\nolimits} ({h_i},h_j^ + )/t}}} }}
\end{equation}

Where $t$ is a temperature coefficient, a hyperparameter controlling the softmax distribution,$sim(h_i,h_j)$ represents the cosine similarity,$N$ represents the size of the batchsize, and $h_{i}^{+}$ represents the augmented sample of $h_{i}$.

(2) KNN-BERT

We added the KNN-BERT method to the model. This method combines a linear classifier with a K-Nearest Neighbors (KNN) algorithm and uses a weighted average as the final prediction score.
\begin{equation}
S = (1 - w){\rm{Softmax}}(F(t)) + w{\rm{KNN}}(t)
\end{equation}
Where $w$ is the weight ratio, $F()$ is the linear classifier, and KNN logits is a voting result, denoted as $\mathrm{KNN}(t)$.
\subsubsection{Matching layers}
The Sentence-BERT model obtains two text vectors U and V, respectively, splices them together, and multiplies them by a trainable weight, using a cross-entropy loss function in the optimization phase. Finally, the semantic relevance score is calculated by the Softmax function. Here we change the loss function to the Cosent \footnote{Retrieved from https://kexue.fm/archives/8847} loss function, a supervised loss function. In the Sentence-BERT model, for the samples marked in the text matching corpus, the goal of the loss function tends to be 1 for the positive samples and 0 for the negative samples. This will make the model lose its generalization ability or be difficult to optimize.

For the above problem, Cosent has improved it by making the similarity of positive sample pairs greater than the similarity of negative sample pairs and introducing the corresponding loss function (4,5).
\begin{equation}
t^{*}(1-\cos (u, v))+(1-t)^{*}(1+\cos (u, v)) \quad \mathrm{t} \in\{0,1\}
\end{equation}
\begin{equation}
\cos \left(u_{i}, u_{j}\right)>\cos \left(u_{k}, u_{l}\right)
\end{equation}
\begin{equation}
{\rm{log}}(1 + \sum\limits_{(i,j) \in {\Omega _{pos}},(k,l) \in {\Omega _{neg}}} {{e^{\lambda (\cos ({u_k},{u_l}) - \cos ({u_i},{u_j}))}}} {\rm{  }})
\end{equation}
Where  $\Omega _{pos}$ is a set of positive sample pairs, $\Omega _{neg}$ is a set of negative sample pairs. $u_i$ and $u_j$ are positive sample sentence vectors.$u_k$ and $u_l$ are negative sample sentence vectors. $\lambda >0$ is a hyperparameter.

\subsection{Training process}

\begin{figure}[!htbp]
  \centering
  \includegraphics[width=13cm,height=6cm]{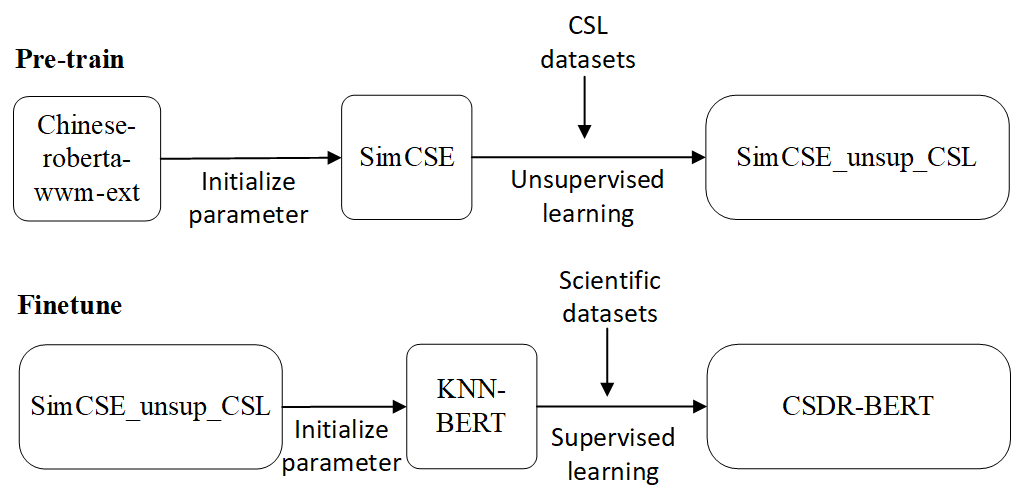}
  \caption{CSDR-BERT training process}
  \label{fig:f3}
\end{figure}

In the pre-training phase, the language model focuses on learning the basic semantics of words and the semantic dependencies between words. In the fine-tuning phase, the model will be enhanced for specific tasks. Therefore we use the SimCSE method for unsupervised learning of the CSL dataset in the pre-training phase, by which we hope to memorize the semantic information of each scientific data in the model. The fine-tuning phase uses the above pre-training model to initialize the parameters and learn the semantic relationship between two sentences via the KNN-BERT method. During this process, the model is continuously optimized through the Cosent loss function, thus improving the model performance. the CSDR-BERT training process is shown in Figure \ref{fig:f3}.

\section{Experiment}
\subsection{Experimental environment and parameters}
In this experiment, we use Linux as the operating system, NVIDIA Tesla V100 as the GPU, and CUDA 10.1 to fine-tune the experiment. -The initial learning rate is 0.01, the minimum learning rate is 0.00002, and the batchsize is 32. In the initial training of the model, we first perform a warm-up training with the dataset length * 0.05 steps. The text vector output is CLS.
\subsection{Datasets}
In this paper, we collected more than 30,000 headings and other information from the National Earth System Science Data Centre (NESDC), pre-processed the data using data cleansing and de-duplication, and manually annotated the data to build similar semantic sentence pairs. In terms of the annotation results, the similarity was divided into two levels: 1 for similarity and 0 for dissimilarity, and a training set and a test set were constructed in the ratio of 8:2 (hereinafter referred to as the "scientific dataset"). The training set consists of 8607 pairs and the test set consists of 2353 pairs.
Chinese Scientific Literature Dataset (CSL), containing meta-information (title, abstract, keywords, subject, discipline) of 396,209 core Chinese journal articles. The data are sourced from the National Engineering Technology Research Centre for Science and Technology Resources Sharing Services. We used a dataset of 10,000 articles from the CSL Benchmark.
\subsection{Evalution metric}
The evaluation metric for the similarity semantic task is to determine whether the semantics of pairs of input sentences are equivalent. We use two evaluation metrics commonly used in this field: F1-Score, Accuracy, and spearman correlation coefficient to evaluate the model performance. The larger the evaluation metric, the better the model performance. The corresponding evaluation metrics and calculation formulas are as follows.

Precision:
\begin{equation}
Precision=\frac{TP}{TP+FP} 
\end{equation}
Where $TP$ means that the sample is true and the prediction is true, $FN$ means that the sample is true and the prediction is false, $FP$ means that the sample is false, $TN$ means that the prediction is true, the sample is false and the prediction is false.	

Recall:
\begin{equation}
Recall=\frac{TP}{TP+FN} 
\end{equation}

F1-score combines accuracy and recall metrics:
\begin{equation}
F1-score=\frac{2*Precision*Recall}{Precision+Recall} 
\end{equation}

Accuracy:
\begin{equation}
Accuracy=\frac{TP+TN}{TP+TN+FP+FN} 
\end{equation}

Spearman correlation coefficient:

\begin{equation}
\rho=1-\frac{6 \sum d_{i}^{2}}{n\left(n^{2}-1\right)}
\end{equation}

\subsection{Baseline}

To confirm the validity of our models, several baseline models were selected for comparison, including the SimCSE pre-training model and the Sentence-BERT text matching model.
SimCSE: It is a comparison learning method. The same sentence of text is split twice through the model, but using two different dropouts, so that the two resulting sentence embeddings are used as positive examples of the model, while the other embeddings in the same batch are used as negative examples.
Sentence-BERT: With the development of large-scale pre-trained language models such as BERT, the Sentence-BERT text matching model was derived, where two sentences are encoded into a representation vector, and the result is used as the classification result by calculating the cosine similarity, etc.

\section{Experiment}
By improving the Sentenc-BERT model, ablation experiments were conducted on the scientific dataset in this paper to demonstrate the effectiveness of our improved model, and the experimental results are shown in Table 1. Where Robert denotes the added Chinese-Robert-wwm-ext pre-training model; Cosent denotes the loss function was modified to the Cosent function; SimCSE denotes the pre-training model was changed to SimCSE-unsup-CSL model; KNN denotes the KNN-BERT module was added. We conducted experiments with 30 epochs, and the experimental results are shown in Table \ref{tab:table1}.
\begin{table}[!htbp]
\caption{Comparison of the results of the ablation experiments of this paper's method on the scientific dataset}
\centering
\begin{tabular}{@{}lll@{}}
\toprule
Approch                          & F-score & Accuracy \\ \midrule
Sentence-BERT                    & 84.61\%                   & 84.47\%                   \\
Sentence-BERT+Roberta            & 86.79\%                   & 86.84\%                   \\
Sentence-BERT+Robert+Cosent      & 88.07\%                   & 88.29\%                   \\
Sentence-BERT+SimCSE+Cosent      & 88.41\%                   & 88.78\%                   \\
Sentence-BERT+Roberta+Cosent+KNN & 92.06\%                   & 92.25\%                   \\
Sentence-BERT+SimCSE+Cosent+KNN  & \textbf{92.49\%}          & \textbf{92.66\%}          \\ 
\bottomrule

\end{tabular}
\label{tab:table1}
\end{table}

As can be seen from Table 1, of the methods proposed in this paper, Accuracy improves by 1.45\% when using the Cosent method, 0.49\% when adding the SimCSE unsupervised model, and 3.88\% when adding KNN. When all the improved methods were added to the model at the same time, the overall Accuracy was improved by 5.82\%, at which point the Accuracy was 92.66\%.
The improved model in this paper was experimentally compared with other mainstream semantic similarity matching models on various Chinese datasets to verify its effectiveness and scalability, and the experimental results are shown in Table \ref{tab:table2}. 
\begin{table}[!htbp]
\centering
\caption{Comparison of experimental results of different models on different Chinese datasets (spearman correlation coefficient)}

\begin{tabular}{@{}lllll@{}}
\toprule
Model                & AFQMC            & LCQMC            & STS-B            & PAWS-X           \\ \midrule
SimCSE-unsup         & 24.68\%          & 70.01\%          & 71.16\%          & 10.26\%          \\
Sentence-BERT        & 43.37\%          & \textbf{79.37\%} & 70.79\%          & 58.90\%          \\
Sentence-BERT+Cosent & 77.84\%          & 78.63\%          & 80.59\%          & 60.73\%          \\
CSDR-BERT            & \textbf{78.26\%} & 78.97\%          & \textbf{80.85\%} & \textbf{62.96\%} \\ \bottomrule
\end{tabular}
\label{tab:table2}
\end{table}

As can be seen from Table 2, the improved model CSDR-BERT in this paper achieves better results than the Sentence-BERT model on most of the Chinese datasets. the unsupervised approach of SimCSE works poorly on the PAWS-X dataset because the samples in this dataset have more stacked words but different semantics, making it impossible for the model to make effective judgments by unsupervised clustering. The essence of Sentence-BERT is to use twin networks to represent the text to be matched as a semantic vector by BERT embedding and to extract the features of the two sentences in different ways for stitching and outputting the results by Softmax. samples and can learn domain knowledge. The experimental results also demonstrate the effectiveness and scalability of our approach.

\section{Discussion and future work}
\subsection{Analysis of the need for deep semantic matching models for scientific datasets in open science}

With the rise and development of the data-intensive scientific research paradigm and data science, the role of scientific data in supporting and safeguarding scientific research, science, and technology innovation has become more and more obvious. At the national level, scientific data and academic literature resources have been categorized as an important part of national infrastructure by developed countries such as Europe and the US, and the EU, the US, and Germany have formulated relevant strategic plans and policies to promote the sharing and reusability of scientific data. In late 2018, the General Office of the State Council. Notice on the Issuance of Measures for the Management of Scientific Data" to further strengthen and regulate the management of scientific data, safeguard the safety of scientific data, improve the level of open sharing, and better support national scientific and technological innovation, economic and social development, and national security. Numerous well-known publishers, grant funding agencies, research institutions, and consortia of societies have formulated scientific data-sharing policies. Publishers have explicitly requested or advised authors to submit relevant supporting data along with their papers and to assign permanent unique identifiers (e.g. DOI) to literature and data respectively, while data journals specializing in publishing scientific data describing papers have also emerged. It also calls for the creation of interconnection mechanisms between publishers and data repositories to facilitate access to and linked discovery of resources such as scholarly literature and scientific data.

\subsection{Analysis of the applicability of pre-training techniques in the retrieval of Chinese scientific datasets}

When the commonly used text matching models were applied to the scientific data domain, we found that these methods performed moderately well in Chinese. After analysis, it was found that the semantic complexity of sentences in the scientific domain contains information models such as metadata that cannot be recognized. In addition to retrieval problems, the inconsistent level of questions from users and the different language styles pose a great challenge to the processing of textual information. Therefore, we need to investigate new methods to address these challenges. We have improved the Sentence-BERT model to better suit the retrieval task by replacing the generic pre-training model of Sentence-BERT with a pre-training model of the scientific domain to better recognize domain information.

\subsection{Future work}

The text matching models we have studied can meet the needs of scientific data retrieval. In addition, we found that pre-training models are missing in the scientific data domain, the small number of publicly available scientific datasets, how to do similarity calculations between metadata and title text, and reordering. In the next work, we will introduce more research, such as the annotation of scientific datasets and pre-trained models in the Chinese scientific data domain, to further optimize the text matching task in the scientific data domain.

\section{Conclusion}

In this paper, we propose a CSDR-BERT model for semantic text matching in the Chinese scientific data domain. The model takes full advantage of the latest deep learning "pre-training + fine-tuning" research paradigm, where the model enhances the text representation through a contrast learning approach, learns domain knowledge through a SimCSE pre-training model for scientific dataset retrieval, and improves the original Sentence-BERT by adding a BERT-KNN at the model layer. The model achieves optimal results on both public datasets of the self-built dataset. The study further validates the effectiveness of applying pre-training techniques to the retrieval of Chinese scientific datasets.

\section{Acknowledgments}
This work was supported by grants from the Natural Science Foundation Project of Ningxia Province,China titled "User-oriented Multi-criteria Relevance Ranking Algorithm and Its Application (2021AAC03205)",Key R\&D Program for Talent Introduction of Ningxia Province China titled "Research on Key Technologies of Scientific Data Retrieval in the Context of Open Science (2022YCZX0009, 61862001)",Starting Project of Scientific Research in the North Minzu University titled "Research of Information Retrieval Model Based on the Decision Process (2020 KYQD37)" and North Minzu University Postgraduate Innovation Project (YCX22178, YCX 22193).

\bibliographystyle{unsrt}  
\bibliography{references}

\end{document}